# Growth of global science: reflections on measurability and scientometric insights

Robert Tijssen[1,2*], Erik van Wijk[1], Alfredo Yegros[1]
1. CWTS, Leiden University, Netherlands
2. CREST, Stellenbosch University, South Africa

* Corresponding author (tijssen@cwts.leidenuniv.nl)

## Abstract

Global science has become much larger and more interconnected in recent years. Such a complex and adaptive social system defies any attempt to systematically and comprehensively measure or monitor its growth and evolution. Yet, as it continues to expand and enhance its public value, effective evidence-based assessment has become increasingly critical for understanding its dynamics and managing the system.

This paper offers a framework for addressing system-level measurability challenges. Inspired by Derek de Solla Price's pioneering scientometric work in the 1960s, this paper addresses analytical issues related to current measurements of the system's expansion since the turn of the century. Our empirical analysis of global science dynamics presents new macro-level data on several observable 'megatrends' since the year 2000.

Acknowledging the limited analytical scope of these trend data to understand the dynamics of global science, we outline currently available options to improve system-level measurability under conditions of ambiguity and uncertainty. Looking forward, we discuss likely future directions to establish integrated, multi-stakeholder measurement infrastructures.

## Keywords
History of contemporary science; Scientometrics; Derek de Solla Price; Global trends; Measurement and monitoring

# 1 Measuring structural changes in global science

## 1.1 In the footsteps of Derek de Solla Price

More than 60 years ago, Professor Derek de Solla Price (DdSP) observed an exponential growth of science in the 19th and 20th century and assumed, by extrapolation, that science would inevitably hit a resources ceiling in the 21st century.[1] In his seminal and highly-cited book entitled '*Little science, Big Science'* (De Solla Price, 1963), he developed a quantitative historical approach to explore and describe the nature and driving forces of that expansion. In the posthumously published *Little Science, Big Science … and Beyond*, he states (De Solla Price, 1986, p. 17):

> "*Scientists and engineers are now a couple of percent of the labor force of the United States, and the annual expenditure on research and development is about the same fraction of the Gross National Product. It is clear that we cannot go up another two orders of magnitude as we have climbed the last five. If we did, we should have two scientists for every man, woman, child, and dog in the population, and we should spend on them twice as much money as we had. Scientific doomsday is therefore less than a century distant*".

According to DdSP's predictions, global science could not continue to grow at the same rate for several reasons: it would require major financial investment, concerted efforts and policies of funders and, above all, large numbers of talented individuals and experts who, working under the right conditions, would keep producing new ideas, discoveries and scientific breakthroughs. DdSP discusses the evolution of science from small-scale individual endeavors (which he labelled 'little science') to large-scale, collaborative initiatives ('big science'). This gradual transformation has been shaped and fueled by rapid accumulation of scientific knowledge and technological advances. While DdSP initially focused on natural sciences and engineering sciences, similar trends and trajectories have also been observed in the academic development of sociology (Perrucci et al., 2017).

DdSP's studies constitute one of the earliest empirical accounts of large-scale changes and transformations in global science, a topic that has attracted several high-profile accounts of modern-day transformations (e.g. Adams, 2013; Wang and Barabàsi, 2021). He was also one of the early adopters of large-scale scientometrics, back then an emerging branch of information science that started in the 1920s. This paper follows in DdSP's footsteps by applying scientometrics to study macrolevel trends in the 21st century global science ' system'.[2] As such, it fits within a small but rapidly growing body of scholarly work, partially based on Research and Development (R&D) statistics and other quantitative data, devoted to monitoring and understanding the structural changes that science and its research-intensive universities has been undergoing since the late 20th century (e.g., Fernandez-Cano et al., 2004; Milojević, 2015; Tijssen and Winnink, 2016; Marginson, 2021; Adams and Szomszor, 2022; Baker and Powell, 2024; Kang et al., 2024).

---

[1] The general concept 'science' is loosely defined here as "systematic production and organization of information and knowledge that can be rationally explained and reliably applied".

[2] The science system is a dynamic social system - in contrast to physical system like the weather. This social and complex system is partially self-organizing where its constituent elements, processes and outcomes are heavily affected by both within-system and outside-system factors such as psychological, social, cultural, political and economic factors and circumstances. This open system comprises of multi-varied, interconnected balances between supply and needs for scientific information, knowledge, skills, expertise and facilities, where underpinning scientific activities are both collaborative and competitive.

### 1.2 Science is a complex, dynamic social system

Has global science become too large and complex for advanced measurement and monitoring? Since DdSP published '*Little Science, Big Science*' in the 1960s, almost 80 million new scientific papers have been published in international scientific journals, alongside approximately 15 million patents.[3] What DdSP didn't consider in his predictive model of science's modern-day expansion was the notion of science as a 'non-linear' adaptive system. Changes in recent decades bear witness to the fact that the system's self-organizing and dynamic nature has the capacity to introduce discoveries and radically new technical facilities, such as the internet, which have dramatically changed its idea-generating mechanisms and knowledge production processes.

Science was and is largely driven by human agency and can therefore best be viewed and described as a highly dynamic social system. This 'complex' system is characterized by a stunning array of processes, interactions and feedback loops among numerous actors and contributing factors, constant change, and deeply embedded social structures and norms. It is a partially open system, with many intricate connections to entities outside science and comprising of many ill-defined and overlapping knowledge domains.[4] Owing to its complexity and ability to self-organize and adapt to external and internal changes, its evolution and future development is unpredictable. The elusive and multifaceted nature of this complex system requires sophisticated models and analytical frameworks to understand and monitor system-level developments and changes (Zeng et al., 2017).

### 1.3 Research issues and questions

Measurement, in the broad sense, is defined as a process of empirical, objective assignment of symbols to attributes of objects and events of the real world, in such a way as to describe them (Finkelstein, 2003; 2009). Measurements are tools of information reduction. This process can be 'strongly' defined, where measurement conforms to the strict paradigms and measurement instruments of the physical and natural sciences. In the case of the global science system, and social sciences analytical toolkit than we can bring to bear, in most cases measurements are 'weak' and produce data of sometimes questionable validity and relevance for describing features of that system. DdSP was acutely aware of the difference between strong and weak measurements, judging by this statement (De Solla Price, 1986, p. 139):

---

[3] The patent output refers to the number of applied patents in the USPTO system, the US patent office that was established prior to 1963 (other major offices started in a later year). The research publication output count comprises the *Web of Science Core Collection* databases for the years 1963-2023.

[4] Reflecting its interactive, human-driven nature, one could refer to global science as an 'ecosystem', but also as a 'global sciences' system' (plural) because of its heterogeneity in terms cognitive and epistemic diversity. For the sake of simplicity, we will retain the more generic, singular and neutral term 'global science system'.

*"A word of advice is necessary, especially to social scientists, about the way in which these numbers are to be used. I suggest that they are not to be taken merely as empirical data to be correlated and projected in the usual way. I think that, instead of the usual cookbook statistical methods, the paradigm should be the methodology of astrophysics in the days of Jeans and Eddington. One took whatever data came from the heavens, since experiment and measurement were out of the question, and then one tried to find suitable laws and models that would explain why these observations had the form and the order of magnitude that they did.*
*Thus, if it is stated that a country spends 4 percent of its gross national product (GNP) on Research and Development (R&D), or if one says that Israel has 1 percent of the world's population of nuclear physicists, or that one-fourth of all scientific discoveries are rediscoveries, we can only admit that such a statement is meaningful if we know what we should expect and why. One does not require the accuracy and technique used in testing Boyle's Law, nor the laboriousness of regressions and correlations; one needs the same rough and ready methods, initially, as in asking how many stars there should be in a galaxy."*

Even such 'rough and ready' methods are subject to critical reflection and quality control when applied in scientific research. The concepts of reliability, validity, generalizability and verification are therefore central to weak measurement. On the issue of validity and verifiability DdSP notes (De Solla Price, 1986, p. 3):
*"… we must begin our analysis of science by taking measurements, and in this case it is even more difficult than usual to make such determinations and find out what they mean."*

The issue of assigning meaning to measurements requires an investigation of measurability. Any weak measurement is, by definition, a gross simplification of reality; a reductionist tool that inevitably suffers from measurement limitations and methodological shortcomings that may undermine its legitimacy as an acceptable representation of global sciences. The selected metrics are largely shaped by the limited availability of internationally comparable high-quality information and will disproportionately value what is comparably easy to measure. So, unavoidably, the model introduces incongruences between its data-constrained analytical framework and complex, real-life realities of science communities and their environments.

Although the application of the 'system' metaphor inevitably leads to problems of reductionism and functionalism (e.g. Papaioannou et al., 2009), it also provides an apt conceptual model for exploring and discussing the limits and options of science system analytics and measurability. Measuring the nature and dynamics within large and complex social systems like the scientific community at a global level presents a range of challenges: while many factors can be measured to a significant degree at lower levels of aggregation, even in those cases there are inherent challenges, particularly in areas requiring qualitative assessment or where data is unevenly distributed across countries, regions and fields of science. With this backdrop of constraints and opportunities, this paper will address the following three research questions on the information value and limitations of model-based measurability for system-level mapping and monitoring:

- how can we monitor major macro-level developments in the current global science system with reliable empirical data?
- to what degree do current measurement models and information sources produce relevant and valid information on large-scale changes occurring within that system?
- what can we learn about current measurement limitations to improve system-level measurability and monitoring?

The next section deals with the first of these three questions by presenting annual 'research and development' data that describe and track of some major trends in the 21st century science system. Section 3 addresses the second question, while section 4 deals with the last question and concludes with general observations and recommendations on possible future options.

# 2 Views of 21st century R&D trends in the science system

## 2.1 Analytics and information sources

The global science system has seen significant transformations in the 21st century, driven by advancements in technology, by globalization as well as shifts in funding priorities and policy frameworks. Tracking changes over time to identify long-term R&D trends requires suitable empirical data and an analytical 'science of science' framework to synthesize these data sources and interpret observable trends effectively (Liu et al., 2023). As for the data sources, one of the most accessible sources of empirical information is bibliometric analysis, which extracts its information from R&D-based publications such as research articles and patents. At the current stage of technological development, the 'bibliometric analysis' approach is in fact the only one available that meets crucial methodological requirements that ensure measurements with a fair degree of validity and relevance (Sugimoto and Lariviere, 2018). To measure aspects of scientific research and monitor macro-level developments in the global science system, bibliometric analysis focuses on analysing publication data and citation patterns between research publications. In other words, the results of science are shared and utilized within the scientific community.

Focusing on the bibliometrics-based measurement monitoring, this paper explores the boundaries of system-level measurability. By extracting our 'metadata' on scientific progress from research publications, we follow in the DdSP's footsteps (De Solla Price, 1986, p. 138/139):

> "*Because of the constancy in the level of papers, most people concerned with the measurement of scientific outputs have been reduced to using counts of paper, the men who write them, and the discoveries contained in them. The accepted customary technique is to point to the availability of great masses of such data collected for other purposes (such as scientific bibliography) and to the attractiveness of using these by-products of scientific activity as a noninterfering and silent way of measurement. One then points out that it is inevitably much more convenient to make a measurement, which is relatively easy, than to determine exactly what it is that has been measured. After this, one uses the numbers as one pleases.*"

In addition to having such numbers at one's disposal, the other methodological advantage of bibliometrics is availability of large bibliographic databases such as *Web of Science* (WoS),

*OpenAlex*, *Dimensions* or *Scopus* with millions of these publications, as well as internationally comparable metadata on these publications, their interconnections and external linkages. These large-scale bibliometric databases are made available for online searching and analysis by commercial providers (Clarivate and Elsevier) as well as several 'open access' non-profit sources by public entities such as *SciSciNet* (Lin et al., 2023).
Each of these databases cover all fields of mainstream scientific research and, as such, represent a major segment of the global science system. This case study extracts its metadata from the WoS, as available with the CWTS in-house information system. The WoS-based count data comprises all WoS-indexed publications of the 'research article', 'review article', 'letter' and 'conference proceedings paper' document types.

The global output of research publications has expanded enormously. Looking back to the 1950s, Compound Average Growth Rate (CARG) WoS-indexed publication output is about 5% (Bornmann et al., 2021).[5] Other studies show output growth rates of around 8-9% since the 1980s (Bornmann and Mutz 2015; Fortunato et al., 2018). We will apply a range of WoS-based bibliometric indicators to track annual trends from the year 2000, up to 2021; the research publication production of WoS-indexed publications grew by a CARG of 5.5% during that period.[6]

Our 'big data' monitoring includes two measures of research publication productivity: the relative outputs per researcher and per unit expenditure. The underpinning scientometric data are: (i) number of researchers in the Higher Education sector are counted in full-time equivalents (fte) and (ii) the data on 'R&D expenditure in Higher Education' (HERD) are extracted from the OECD's online databases.[7] HERD data used as a macrolevel indicator of longer-term investments in academic scientific research and associated knowledge production and dissemination. These data comprise 47 countries, OECD's member states and OECD-affiliated countries, which includes most of the world's leading science nations and collectively represent more than 90 percent of the global research publications. Data on research funding is essential for understanding general trends and priority shifts in resource allocation.

---

[5] The Compound Annual Growth Rate represents the mean annualized growth rate for compounding values over a given period. CAGR is equal to the average annual increase (or decrease) compounded each year. This metric reduces the volatility of periodic values that can render arithmetic averages less meaningful.

[6] Recent studies show that the publication output decreased by 5% in 2023, as compared to 2022, the result of the Covid-19 pandemic (WIPO, 2024). We assume that the 'traditional' 5-8% growth rate will re-continue once the negative effects of the pandemic on research productively wear off.

[7] The OECD statistics are extracted from its MSTI database. HERD represents the component of a country's Gross Expenditure on R&D (GERD) incurred by units belonging to the Higher education sector (OECD, 2015). It is the measure of intramural R&D expenditure within the Higher education sector during a specific period. HERD includes current and capital expenditure (labor and other current costs), where R&D labor costs should be linked to the time spent on R&D, the full-time equivalent (FTE) measure. Current costs include items such as documents, minor equipment, subscriptions to scientific journals, travelling costs, etc. Expenditures on fixed assets used for R&D, such as machinery/equipment and land/buildings, including real estate and facility management costs. The R&D expenditure data are in real value, i.e. corrected for a country's purchasing power.
In many OECD countries, the natural sciences and engineering fields receive the lion's share of HERD funding. National statistics agencies of UNESCO's Institute for Statistics often provide detailed breakdowns of HERD by field of science.
Owing to the usual delay between R&D funding and the associated production of research publications, we apply a time-lag in the computation of the productivity measure. Following common practices, we apply a 3-year time-lag. For example, the publication productivity level in the year 2021 is based on the 2018 HERD investments.

Focusing the empirical analysis on significant increases in WoS-indexed global research output, which we assume to partially reflect 'growth' of the science system, this metrics-based analysis presents three analytical angles to highlight trends that are driving growth: the rise of China; increased research cooperation; accessibility of research publications.

## 2.2 Observed empirical patterns and trends

**Figure 1** compares the annual growth pattern of publication output, alongside the annual trends in HERD and the number of researchers. Global science continues to expand vigorously in the 21st century. These output growth rates are obviously affected by the increase of human and financial inputs, but also by many contributing factors (see Table 1) and their interplay.

Figure 1: Relative growth of research resources and publication output (index 2000 = 100)

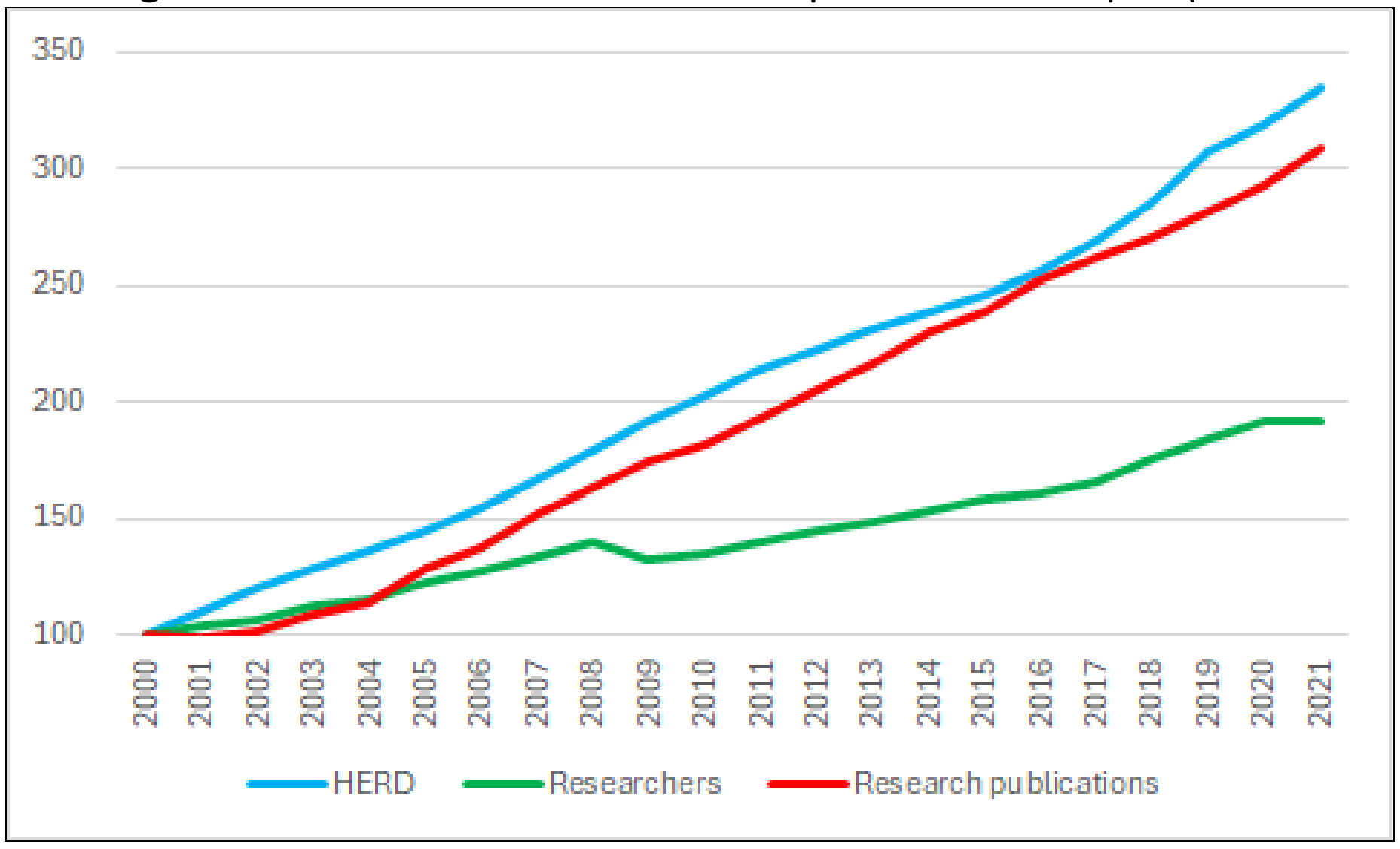

Interestingly, threefold growth rate of publication output is at a similar level as the HERD increase but outpaces the twofold growth of researchers in the OECD countries. Although not all HERD is devoted to research-related investments and costs,[8] the similarity of the HERD and publication output trendlines is explained by the fact that a large majority of WoS-indexed publications are produced by research-intensive universities and other research-active Higher Education Institutions (HEIs). Recent research shows that the expanding HE sector, where funding for education cross-subsidizes research in HEIs, is one of the major contributing factors of growth in global science (Baker and Powell, 2024). Not all HERD is spent on hiring research staff, the growth of which lags significantly. The slowly expanding community of researchers nonetheless manages to sustain an impressive publication output growth rate since the turn of the century.

[8] HERD is the measure of intramural R&D expenditures within the Higher education sector during a specific period. In many OECD countries, the natural sciences and engineering fields receive the lion's share of HERD funding (national statistics agencies of UNESCO's Institute for Statistics often provide detailed breakdowns of HERD by field of science). HERD includes current and capital expenditure (labor and other current costs). Current costs include items such as documents, minor equipment, subscriptions to scientific journals, travelling costs, etc. Expenditures on fixed assets used for R&D (Research and Development), such as machinery, equipment and land as well as buildings, including real estate and facility management costs.

Research output has clearly increased very significantly. To gauge the general trend since the turn of the century, we computed two output/input ratio's: (a) research publication counts divided by HERD level, and (b) research publication output by number of researchers. **Figure 2** reveals that the number of publications per unit HERD expenditure has fluctuated since 2003-2004 around 9.5 publications on average. WoS-based productivity has increased significantly: the average number of publications per researcher went up from 2.0 in 2000 to 3.3 in 2021. However, productivity growth has levelled off since 2015.

Figure 2: Research productivity trends (publication output relative to input)

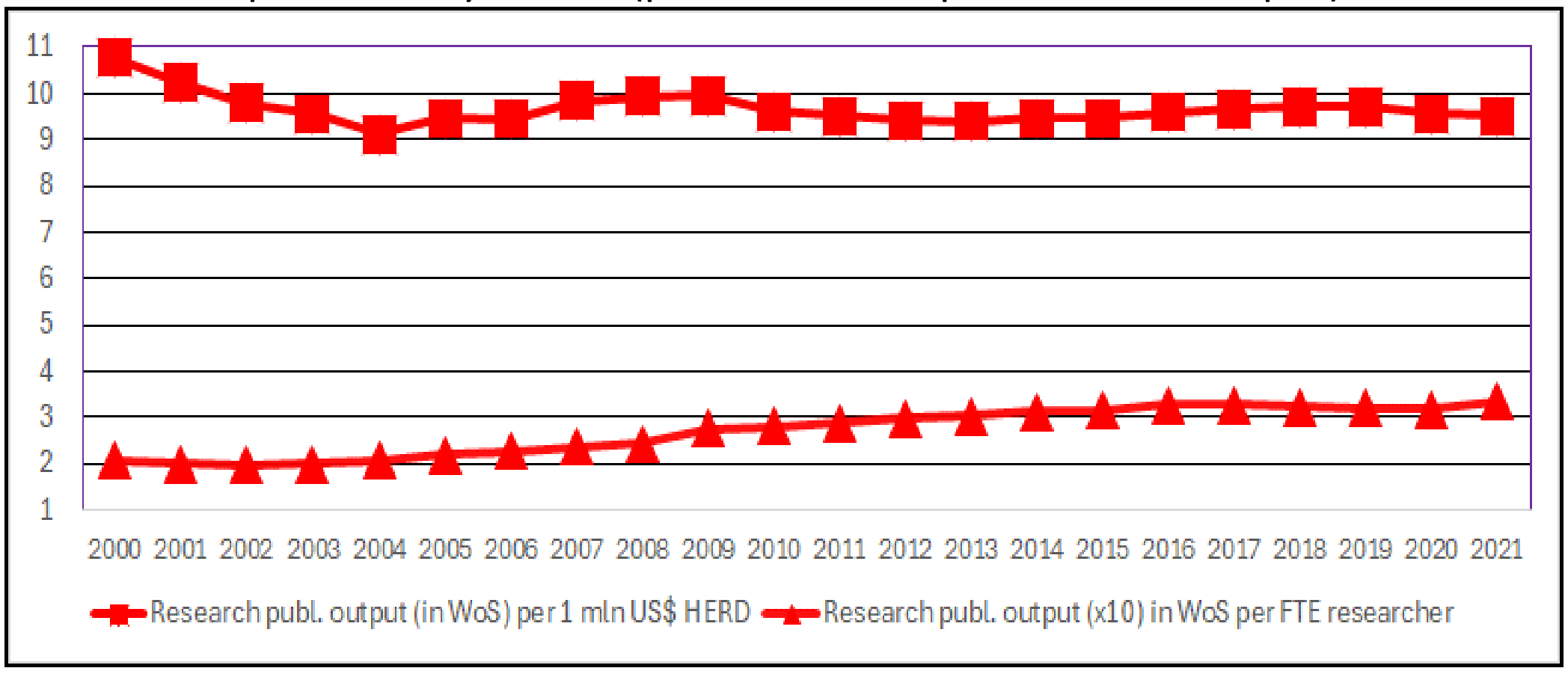

* Geographic coverage: OECD nations (HERD, researchers); all nations worldwide (research publications)

The 65% overall increase of publication productivity within two decades may have resulted from several, mutually reinforcing, factors such as more effective work methods, more opportunities and incentives for engaging in research cooperation, but also more options and available outlets for dissemination research publications. In combination, these two productivity trends suggest that the 'basic science' component within Higher Education R&D activities has become much more productive over time. However, the lack of productivity growth in recent years (despite many advances in digitalization and communication) invites different interpretations such as: (a) the presence of persistent inefficiencies in research processes or information dissemination; (b) the 'overproduction' of publications – owing to 'publish or perish' pressures in some countries such as China - is levelling off because this incentive is considered an increasingly problematic practice in terms of performance accountability and legitimizing the societal relevance of science (Schneegans et al. 2021). As the volume of research publications continues to rise and the burden of peer review falls unevenly across the research enterprise (e.g. Kovanis et al., 2016; Forrester, 2023), the stagnation in productivity rates seems to imply that the science system is under great pressure.

Academic studies have revealed that modern-day science is increasingly driven by cooperation and teamwork (Wagner and Leydesdorff 2005; Gazni et al., 2012; Marginson 2020; Aksnes and Sivertsen, 2023). Adams and Szomszor (2002) find that the rise in national output was strongly associated with bilateral collaboration until the 2000s, but then after that by multilateral partnerships. This shift happens at much the same time across all countries. Studies have repeatedly shown that productive social interactions are one of the drivers of growth rates of research outputs. However, the global research publication rate has not increased when computed at the level of individual authors (Fanelli and Larivière, 2016) where incentive structures might have reshaped publication practices (Fanelli, 2020).

Some of those megatrends in 'global networked science' are exhibited in **Figure 3**. The number of research publications that signify institutional cooperation – i.e. papers with author affiliate addresses referring to different organizations - has increased very significantly. Internationally co-authored publications with research partners based in other countries have increased rapidly in this century (CAGR of 9%). Co-publications with domestic partners expanded at the same annual growth rate. Research collaboration between public sector researchers in universities and research institutes and private sector partners, represented by the global output of public-private co-publications has grown at a slightly lower rate (CAGR of 7%).

Figure 3: Share of institutionally co-authored publications in worldwide publication output

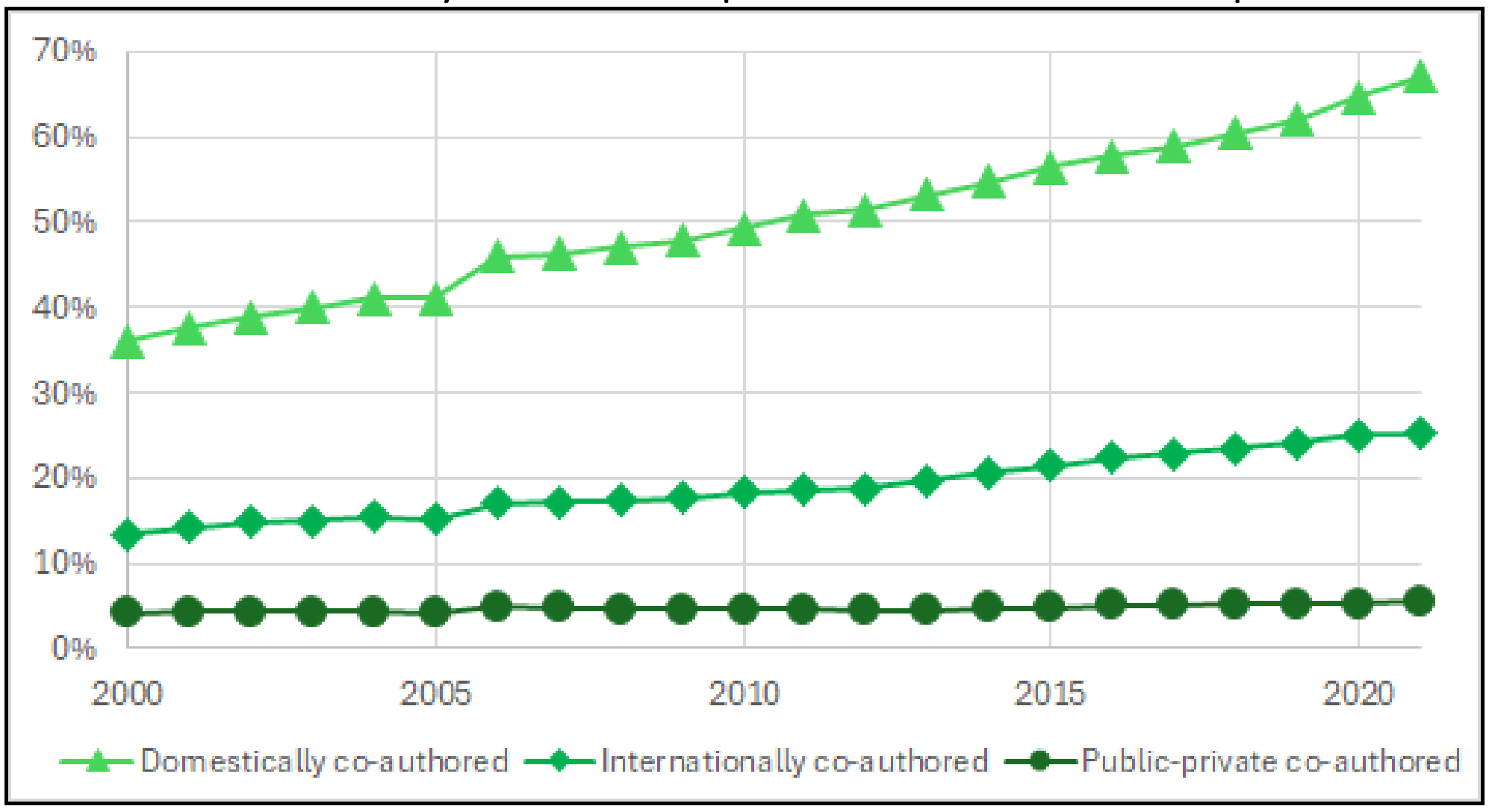

Figure 4: Share of Open Access publications and citedness of all publications

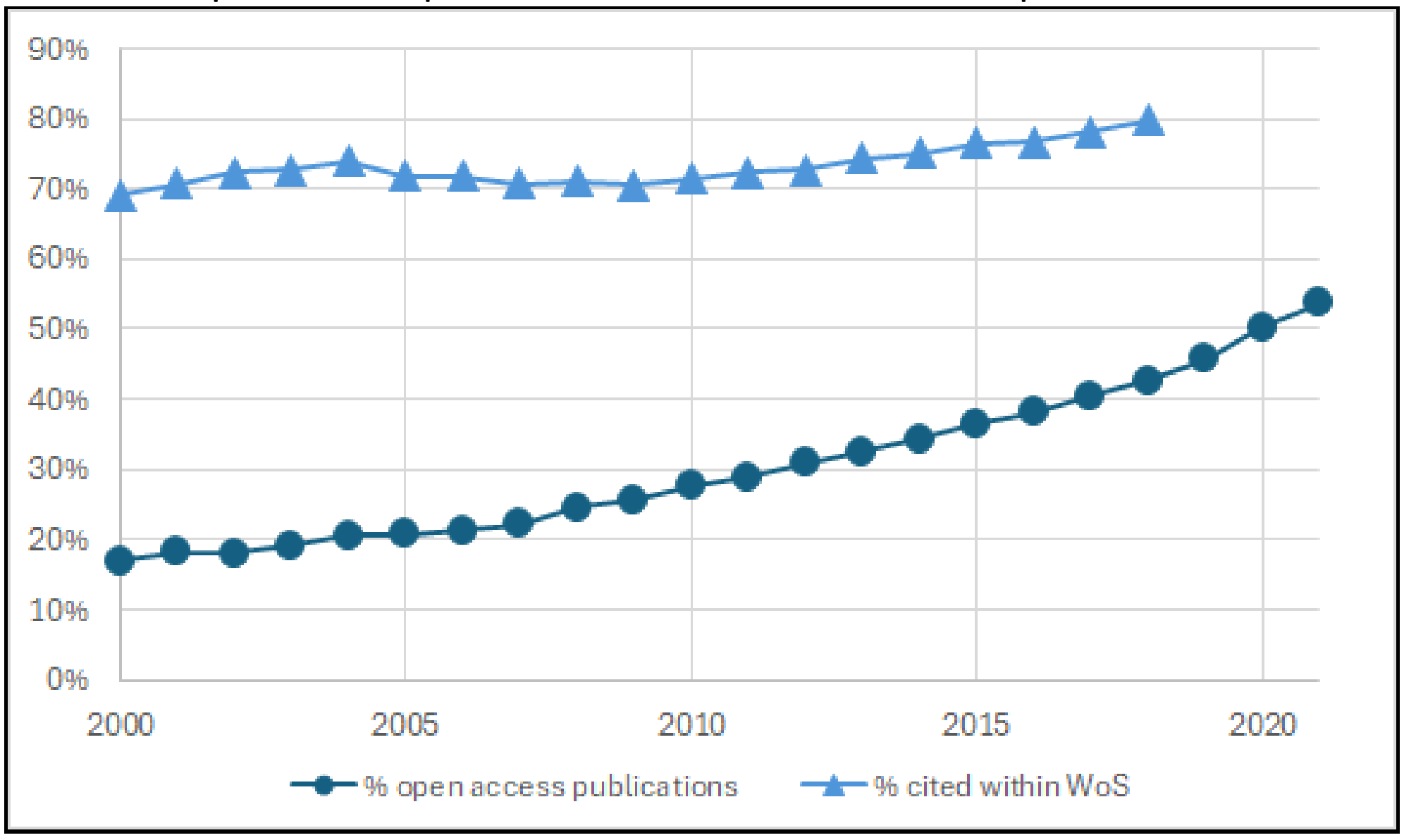

Science's evolutionary change and its ongoing structural transformation processes in communication practices are also manifest in bibliometric trend data. As shown in **Figure 4**, the principles of open science, including open access publication, data sharing, and collaborative research networks, have gained momentum. The share of Open Access (OA) publications in WoS has gone up from 20% to more than 50% in 2021, with a CAGR of 12%. By promoting transparency and accessibility, open science initiatives facilitate the rapid dissemination and uptake of scientific knowledge. Indeed, the share of research publications receiving citations has also steadily increased – by a CAGR of 6% - over the last 20 years, now at 80% of all publications[9]. Citation traffic is also known to be boosted by social connectivity and scientific collaboration between researchers who become more aware of each other's work and research publications.

When examining 21st century megatrends, one cannot avoid paying due attention to the spectacular rise of China that has become a major force in global science during the last 20 years. The country's HERD-level has increased by 14.9% since 2000, the total number of researchers by 6%.[10] This expansion has significantly changed research productivity levels worldwide. China's publication production captured by the WoS database has increased with a CAGR of 15% between 2000 and 2021.[11] The total research publication output expanding rapidly from less than 3% to 18% of the world output. The country contributed a 36% share of the global publication growth in that time-period. China has become such a major force; it is transforming the structure and dynamics of global science. The rise of China and, more recently, an emerging process of 'deglobalisation' and 'decoupling' between China and major Western partners in the USA and Europe may significantly affect the dynamics (García-Herrero and Tan, 2022; Jia et al. 2024).

Figure 5: Chinese share of global publication outputs and cited publications

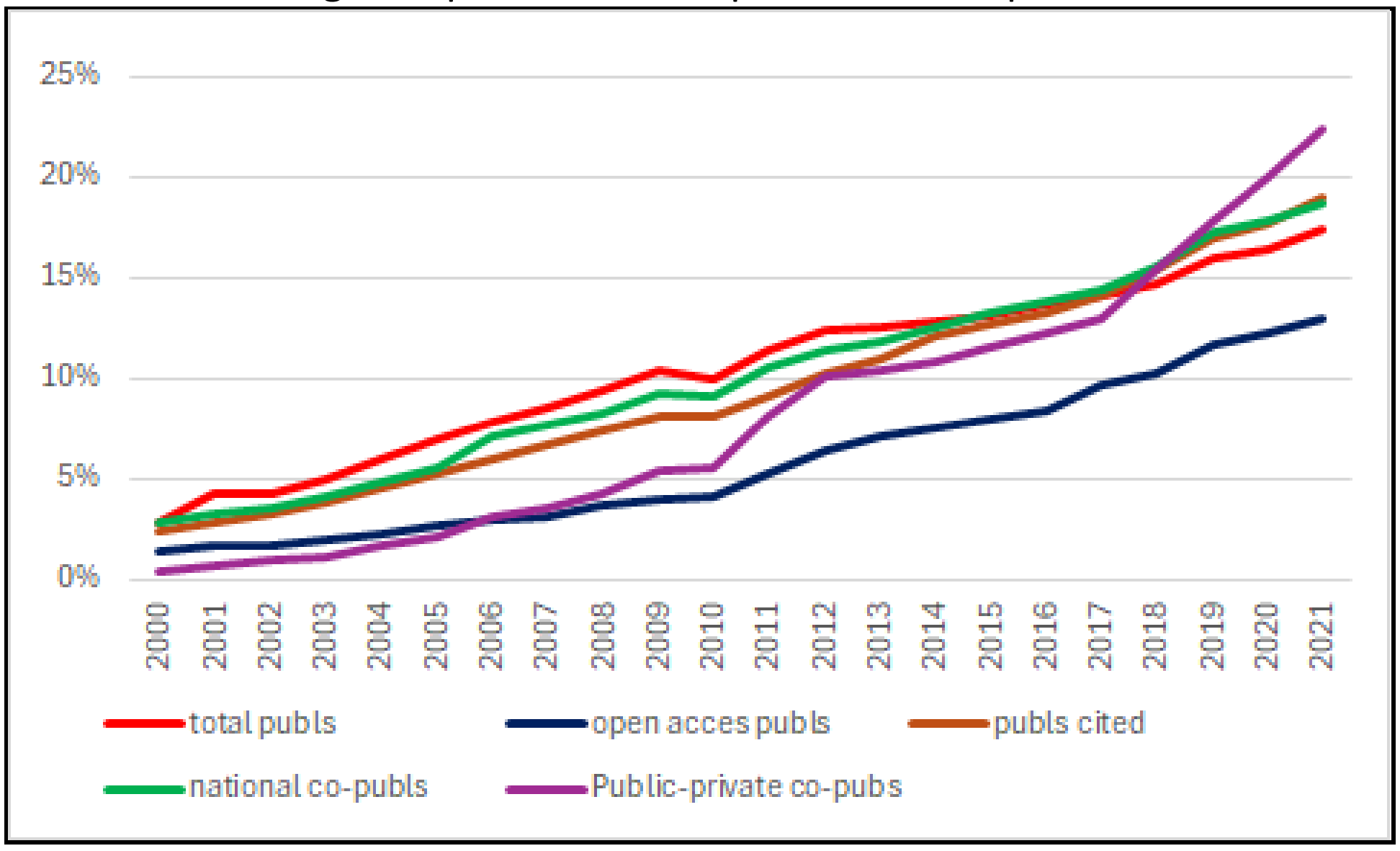

[9] We used a 5-year citation window to gather citation data. The citations to 2018 publications cover the period 2018-2022. Author self-citations are excluded from the computation.

[10] The CAGR scores for the USA are 5.1% (HERD) and 2.5% (researchers).

[11] 'Chinese publications' are defined as those with author affiliation addresses exclusively located in mainland China.

**Figure 5** depicts the steady increase in the shares of China in global research according to several bibliometric indicators. The co-authored research publications involving Chinese researchers from two or more different main organizations in China – 'national co-publications – outpace the overall growth rate with a CARG of 19%. China's share of 'public-private co-publications', i.e. publications involving public sector contributors and their private sector partners, shows an even steeper increase: from 0.5% to 23% (publication output CAGR of 7%). However, China's share of open access publications has increased with a CAGR of no less than 24%.

## 2.3 Interpreting megatrends

DdSP would be surprised by these results and puzzled that modern-day science continues to expand so vigorously. He has been proved wrong - according to the above graphs, global science system is still expanding. Moreover, the exhibited development trajectories suggest further expansion in the years ahead. The graphs show that the continued rise in publication production is closely linked to increases in both the amount of R&D funding and in the number of researchers who are educated in science and trained in scientific work and academic careers (Figure 1). Moreover, a recent empirical study, covering the period 1800-2020 and based on a much larger database of research publications, questions DdSP's observed exponential growth (Kang et al., 2024) – these latest findings display a linear growth pattern of publication output and an exponential growth of the citation links between publications.

DdSP also underestimated the increased importance of research-intensive universities in global science. The massification of higher education in many countries, with concomitant increases of research budgets for old and newly established universities, is seen as another major driving force (Baker and Powell, 2024). The emergence of China as a major player in science (figure 5) has also fueled continued growth during the last two decades. The macro-level trends in Figure 3 also reveal that the system's organizational structure is becoming ever more based on research cooperation. Science is also becoming ever more interconnected – as reflected by citations between publications and research cooperation within and across national and institutional borders (figures 3 and 4). Some say the system has entered the stage of global 'mega science', marked by large-scale research cooperation and continued growth (Baker and Powell, 2024).

But interestingly, research publication productivity has hardly improved in recent decades (see Figure 2) despite better communication and collaboration practices and ongoing digital revolution. This paradox suggests the presence of 'hidden' factors, defying large-scale measurement and monitoring, which are hampering a more effective and efficient way of producing and disseminating new ideas and knowledge. The data presented in Figures 1 to 5 present only a small window on what is happening in the science system and those measurements are not always 'robust'.[12] In the absence of further system-level data one can only speculate about the nature, pervasiveness and impacts of underlying factors and inhibiting obstacles.[13] Hence, owing to the restrictions for the current analytical frameworks, with their limited scope for quantification and comparative measurement, any 'strong' statement with regards to growth (or decline) based on 'weak' metrics is methodologically questionable at the very least and therefore subject to extensive scrutiny.

---

[12] Many bibliometric statistics, such as a 'research productivity score', are also not statistically 'robust' because such statistics can be sensitive to outliers or other departures from statistical distributions. Moreover, the macro-level trends depicted in the graphs reflect time-delayed changes; they are not adequate early warning indicators of emerging changes in the dynamics of global science at lower levels of aggregation. Conversely,

# 3 Confronting complexities

## 3.1 Measurable and non-measurable factors and characteristics

Strong statements require strong empirical evidence that is derived from information sources covering a much broader range of the system's driving factors. Preferably information that is sufficiently amenable to measurement of the system's dynamics and suitable for developing metrics-based proxies ('indicators') of the system's processes, outcomes and impacts. Such a comprehensive measurement model requires a pre-determined organizing framework, including an underpinning theoretical and conceptual model of the science system as well as an overarching analytical model to interpret those measurements. Such a framework will contain a structured overview and a 'classification system' of major components of the science system, its identifiable entities, and empirical information sources to describe them. This theory-laden, analytics-driven model may then enable the computation or estimation of 'system parameter' values based on statistically robust measurement processes.[14]

In this paper, the organizing principle is based on a configuration of abstract interacting spheres of human activity or influence, i.e. 'enabling or driving factors' ('factors' for short) to capture empirical observable properties of the social system. Depending on the factor's scope, characteristics and operational functions, several 'analytical perspectives' can be applied to collect information for measurements. For example, gathering quantitative data on research publication output according to the 'ideas and knowledge production' perspective within the 'cognitive' factor. Lower-level measurements within an analytical perspective may address specific elements, such as for example research publication output of universities (meso level) or that of individual academic researchers (micro level).

**Table 1** presents a non-exhaustive classification of those driving factors and related analytical perspectives. The list of distinctive categories ('factors') in this classification system is primarily designed to help identify information gaps in system-level monitoring

---

significantly recent changes at lower levels might take considerable time before they become noticeable and measurable at the system-level.

[13] Changes in social dynamics are a possible explanatory factor: researchers are under mounting pressure to publish frequently to secure funding, career advancement, and peer recognition. Surviving and thriving in resource-constrained, output-driven and rapidly internationalizing social systems intensifies the competitive atmosphere among researchers and within scientific communities. The expectation to produce more publications at a faster pace may lead to a superficial accumulation of low-quality knowledge rather than major contributions of societal or economic relevance (Schneegans et al., 2021).

[14] Such measurements may then not only describe system parameters, but also explain patterns and processes, and perhaps even predict key processes in the system (thus representing a modern-day model akin to DdSP's exponential growth model from the 1960s to predict the system's expansion rate in the 2nd half of the 20th century).

and measurement. Our succinct labelling of factors and perspectives is primarily meant to introduce a crude categorization where entries are ranked by degree of measurability. Each entry includes our assessment as to the general level of measurability and the type of available information amenable for some types of weak measurement. That subjective assessment is based on our general impressions and a cursory scan of the 'science of science' research literature. The overview in this table is arbitrary in the sense that other factors and perspectives could have been selected according to a different set of epistemological considerations or analytical criteria. Many of the selected factors are closely related and interconnected in a 'non-linear' fashion with many feedback processes and interdependencies. Science, being a complex social system, may react to any of these factors, but not necessarily in predictable one-directional 'linear' way.

None of these entries in this table can boast on a 'very high' level of measurability, but some of the 'high level' or 'medium level' driving forces can be partially described by measurements (albeit still by 'weak' measurement methods). At a lower aggregated level, some elements within these factors – such as the number of researchers or research publication counts – are as highly measurable. Returning to the graphs and their current 'key performance indicators' as displayed in the previous section, Table 1 now reveals how limited the analytical scope of those metrics statistics indicators, and which relevant information on the system's workings is sorely missing in the previous section. The currently available data captures only four of the 27 factors presented in this table. Clearly, numerous important determinants, inputs and underlying processes elude system-level quantification and measurement. These unknowns can lead to misguided policies, underfunded areas of research, and missed opportunities for addressing global challenges effectively. They should not remain invisible and hidden from external scrutiny if we are to develop a more comprehensive overview of the system and its dynamics. Any concerted attempt to address this fundamental issue and its methodological challenges requires dedicated efforts and ample resources to enhance the availability and integration of multiple information sources within and across many factors. The overview of information sources in **Table 1** introduces several options, other than bibliometrics, to gather empirical data.

Table 1: Classification of system-level driving forces

| Driving force | Analytical perspective | Measurability | Main sources of empirical information and data |
|---|---|---|---|
| Resources | people, funds and expenditure, physical infrastructure | high/medium | government statistics; research publications and patents; funding and expenditure data; surveys |
| Behavioural | collaboration and competition | medium | research publications and patents; surveys |
| Behavioural | communication practices, publishing and outreach | medium | research publications; social media; surveys |
| Cognitive | ideas and knowledge production | medium | research publications; social media; surveys; institutional reports and reviews |
| Cultural | scientific practices and paradigms | medium | research publications; social media; surveys; institutional reports and reviews |
| Economic | resource allocation and utilization | medium | funding and expenditure data; surveys; institutional reports and reviews |
| Structural | collaborative structures and networks | medium | research publications; surveys; institutional reports and reviews |
| Technological | science-based technologies | medium | patents; surveys; institutional reports and reviews |
| Behavioural | creativity and innovation | medium/low | patents and research publications; institutional reports and reviews; surveys and interviews |
| Economic | economic impact of science | medium/low | institutional reports and reviews; surveys and interviews |
| Economic | information markets, commercialisation and intellectual property | medium/low | institutional reports and reviews; surveys and interviews |
| Economic | performance incentives; rewards and renumerations | medium/low | institutional reports and reviews; surveys and interviews |
| Ethical | research integrity and security | medium/low | institutional reports and reviews; surveys and interviews |
| Political | policy and regulation | medium/low | policy documents; surveys |
| Political | accountabiliy and evaluation regimes | medium/low | institutional reports and reviews; surveys and interviews |
| Political | policy impact of science | medium/low | institutional reports and reviews; research publications |
| Societal | public engagement, societal impact of science | medium/low | social media; institutional reports and reviews; surveys and interviews |
| Societal | science education, citizen science | medium/low | global educational databases; surveys and interviews |
| Structural | organisational hierarchies | medium/low | surveys and interviews; research publications |
| Structural | institutional structures | medium/low | institutional reports and reviews; research publications |
| Technological | R&D infrastructures | medium/low | institutional reports and reviews; research publications; patents |
| Technological | innovation diffusion | medium/low | institutional reports and reviews; patents |
| Cognitive | epistemological perspectives. cognitive biases | low | surveys and interviews |
| Cultural | identity, diversity and inclusion | low | surveys and interviews |
| Cultural | academic norms and values | low | surveys and interviews |
| Ethical | ethical concerns and regulatory challenges | low | surveys and interviews |
| Political | legitimacy, authority and control | low | surveys and interviews |

## 3.2 Expanding and improving measurability

Owing to those analytical constraints and database dependencies, the scope of large-scale system-level measurements of scientific progress are hampered by many embarrassing information gaps. Relying on R&D statistics and research publications as analytical lenses cannot adequately reflect or capture large, structural trends in the global science system. Improvements in system-level measurability will, at the very least, require new sources of useful information by way of comprehensive and consolidated global databases. Developing a more diverse mix of sources and a larger range of measurement tools will be challenging. The scope, validity and relevance of the analytical framework require a major upgrade. As for validity and scope, traditional bibliometric methods have been the mainstay of monitoring and evaluating global scientific progress since the 1960s.

Naturally, the ongoing evolution in science and structural transformations in recent decades, are also driven by external pressures not sufficiently captured by bibliometric analysis. The reliance on publication counts and citation linkages can lead to an overemphasis on sheer output rather than the quality of the research or other aspects of the research process and the environment in which the work was done. Major segments of the system, such as unsuccessful science or research published in obscure outlets, and key processes such as writing research proposals and reviewing manuscripts remain hidden from external observation. Similarly, the reliance on the few available country-level R&D statistics at the OECD and other statistical offices restricts measurements to country-defined segments of the system.

Take for example, the pressure on public sector research institutes and universities to generate more socio-economic benefits and impacts (see the 'Economics' force in Table 1) driven by demands from governments, industries, and society at large for solutions to pressing challenges such as climate change, health crises, and technological innovation. The complexity of many contemporary challenges and problems requires cooperation and interdisciplinary approaches that draw insights from multiple scientific disciplines. There has for example been a notable shift in global science towards research aimed at addressing real-world problems, such as the United Nations-initiated *Societal Develop Goals* (United Nations, 2015) but gathering empirical evidence on the key characteristics of that shift, and measuring its rate of change, has proven difficult (Jayabalasingham et al., 2020).

As for relevance, obviously any major noticeable change in the science system is open for various interpretations, depending on the observer's analytical perspective and timeframe. For example, returning to Figures 1 and 2, some academic studies suggest that the increased publication production does not necessarily correspond with improvements in scientific quality or originality (Milojević, 2015; Chu and Evans, 2021). Other studies even go as far as to suggest that the rate of scientific progress is in fact slowing down (Bloom et al., 2020; Boeing et al., 2022). A recent study claims that researchers have steadily been producing fewer innovative or disruptive ideas since the 1980s (Park et al., 2023), but this finding is challenged in the research literature on account of methodological error and measurement bias. Replication studies show either the absence of this decline or at a much lower rate of decline (Holst et al., 2024; Macher et al, 2024). More studies and measurements are needed to create clarity on these issues.

As for Figures 3 and 4, how will the drive towards teamwork, open access and research cooperation impact on the creativity and innovation of individuals? Will the surge of OA publications affect communication practices and the role of commercial publishers? To what degree will OA-committed researchers be forced to share their raw data and methodologies? And how will the increasing number of researchers worldwide affect research quality control systems such as the peer review system?

Shared research infrastructures, large-scale physical instruments and pooled information sources have made scientific research ever more collaborative, thereby expanding the capacity for more discoveries. However, some natural resources in the science system are limited. So, future growth rates will inevitably slow down, level off, and science might even come to a standstill in terms of generating new 'breakthrough' ideas. Of course, science itself will have changed by then, where our current notions of growth, productivity or other properties of science system will also have evolved accordingly.

### 3.3 Global surveys and social media data

Where to begin to address information gaps and missing measurements, especially in the case of the 'hidden' non-measurable factors? The list of 'medium-level measurability' factors in Table 1 offers an entry point and guidance as to where progress is probably easiest to achieve. Running down the list of these entries, it becomes clear that relevant, additional system-level information could be extracted from either global surveys or form social media platforms. Both sources already have established a record as sources of relevant qualitative information and quantitative data: survey data are valuable for evidence-based science policy; social media data are increasing used in science studies.

Global surveys are a powerful analytical tool for tracking and describing changes in global science. Such large-scale surveys gather their data from a wide range of respondents across different countries, disciplines, and sectors, providing valuable insights into the evolving landscape of science and research. Survey-based data and indicators are included in several reports that assess the state of global science system and evolving trends such as UNESCO's series of *Science Reports*, the OECD's *Global Science Forum* Report, the Wellcome Trust's *Global Monitor*, the WIPO's *Global Innovation Index*, as well as reports by other agencies such as the International Science Council and the Global Research Council. Collectively these reports and their surveys offer a wide range of relevant data – usually at the level of countries or world regions - into various aspects of scientific research, governance, international cooperation. As surveys often draw their information from opinions and views of sampled respondents, they are vulnerable to subjectivity, sampling biases as well as the potential for manipulation.[15] Nonetheless, when sufficiently sound methodologies are in

[15] There are many other smaller-scale surveys dealing with important components of the global science system. Notably, the surveys of universities worldwide, distributed by commercial providers of university rankings (specifically, *Times Higher Education* and *Quacquarelli Symonds*) that offer information on academic and

place, these surveys may offer a reliable multi-dimensional view across a variety of stakeholders. Surveys on public attitudes toward science and public understanding of science are particularly relevant for system-level monitoring (e.g., Bauer and Falade, 2021). Overall, well-designed, high-quality surveys can be an indispensable additional source of comprehensive data for tracking and describing changes in global science across many of the factors listed in Table 1. Survey-based data can supplement the range of system-level measurements and performance indicators depicted in Figures 1 to 5 and in doing so offer relevant insights to help explain those megatrends.

Social media metrics, initially introduced as 'Altmetrics' (Priem et al., 2010), taps into communication patterns by researchers and others in the science system. These metrics have emerged as a novel and increasingly important source of information to measure the immediate impact and broader visibility of scientific research through various social platforms, offering a complementary view to traditional bibliometric indicators like citations which can take months or years to reflect research impact. Social media metrics may provide further insights into the behavioural, cultural, and cognitive factors underpinning the trends depicted in Figure 4 by capture real-time engagement with research and its societal relevance.

Applications range from contributions to research performance evaluation processes (e.g. Arroyo-Machado and Torres-Salinas, 2023) to measuring the societal impact of science (e.g., Woolston, 2023). Social media metrics can also help map out and analyze the interactions of online communities of scientists, including how different scientific fields form communities or clusters of discussion and collaboration. And as open-access publishing becomes more common, social media metrics are playing a growing role in assessing the reach and engagement of openly available research. However, despite their analytical potential for gathering addition information on some of the driving factors mentioned in Table 1, there are several challenges and limitations to using social media metrics in studies at the global system level, either for monitoring the diffusion and utilization of scientific information within academic and non-academic audiences or for changes in the public engagement with science.[16]

## 4 Reflections on future directions and developments

The global science system is in perpetual change at many different and interacting levels. Understanding how scientific knowledge is disseminated and utilized in society is complex and inadequately captured by empirical information. Any evidence-supported assessment or

---

employer reputation. These surveys are of a more limited value for system-level monitoring on account of their limited coverage of all research-active higher education institutions worldwide, insufficient transparency about the data collection methodology, and therefore questionable validity and reliability of their outcomes.

[16] Popularity on social media platforms can be driven by sensationalism or the novelty of topics, rather than scientific rigor; social media engagement therefore tends to be short-lived, meaning that a research paper might receive attention initially but fade quickly, unlike citations that accumulate over time and often reflect lasting impact; social media metrics can vary significantly across platforms (such as LinkedIn or Twitter/X), making it difficult to standardize and aggregate them for longer-term monitoring at the global level.

judgement of the system's current dynamics or its future development trajectory is clouded by analytical uncertainty and ambiguity. Owing to a lack of external standards, objectification or quantification is neither straightforward nor unequivocal. However, there is a need for more and better metrics, especially to assess knowledge transfer processes and their societal implications. Adopting a more sophisticated multi-method/multi-source methodology for information gathering would, in principle, enable in-depth tracking of changes in global science where high-quality measurements of clearly identifiable empirical processes may help to gain a better understanding of some dynamics. But given the lack of more comprehensive data, detecting large-scale and deep 'structural' changes within the system is beyond present-day analytical tools and measurement.

The integration and alignment of the methods and sources would require a sophisticated information infrastructure and a dedicated architecture comprising a diverse array of comprehensive, relational and interoperable databases (a 'global data lake') with range of robust analytic tools and indicators to produce compelling empirical evidence. Such an advanced measurement and 'big data' analytics system should also be able to capture both 'linear' and 'non-linear' processes in global science. It should include 'vital signs' key indicators of the system's health but also 'early warning' indicators of impending problems and crises. Such a level of sophistication a way beyond our current technical competencies and capabilities, although future developments in AI are likely to resolve some of these issues. Big data analytics are now able to process and analyze large volumes of scientific data for uncovering system-level patterns and trends that might not be evident through traditional methods, while AI and machine learning algorithms could predict emerging research areas, collaboration networks, and the future impact of current research. AI is already making some in-roads as an information support tool for evidence-based science policy (Tyler et al., 2023).

Making technical and infrastructural progress on measurability means organizing large-scale collaboration, enhancing data collection, standardizing existing metrics, consolidating and expanding databases, and leveraging the benefit of existing advanced information technologies. Some information systems for real-time data collection to monitor ongoing research activities and emerging trends in research publications are already in place at commercial information providers. Next steps would require data sharing and adoption the same quality standards for measuring and data analytics to ensure consistency and comparability across countries and institutions. Such information platforms require feedback mechanisms where researchers and institutions can provide input on data collection processes and metrics, leading to continuous learning and improvement. Such an integrated global approach would provide crucial incentives for researchers, institutions and other stakeholders to adhere to standardized international measurement and reporting practices.

Multi-stakeholder international platforms for integrated monitoring of policy-relevant changes in science, driven by metrics and performance indicators, are however bound to spark debate within the academic world, for instance on risks of misuse and biased interpretations of available data. Early attempts to introduce large-scale measurement and metrics-based indicators, especially for evaluative objectives, were considered controversial and met resistance (e.g., Godin, 2005). Moreover, system-wide comprehensive measurement is likely to raise ethical objections about the transparency of micro-level data as well as concerns regarding data manipulation mechanisms, institutional power dynamics and vested strategic interests. Different stakeholders tend to interpret data through varying lenses, complicating consensus. Because most science funding is distributed at the national level, rather than internationally, it may expect resistance from national science policy circles. In conclusion, an internationally authoritative '*Global Science Monitor*' is unlikely to emerge any time soon in view of practical, ethical and technical complexities involved.

Expanding on the existing information sources (bibliometrics, surveys or otherwise), concerted efforts should focus on strengthening policy frameworks that support the expansion and alignment of measurement, monitoring and data-informed evaluation of scientific activities. Major actors in global science and large funding agencies should step in and provide international support and seed funding. Major contributors and beneficiaries of the science system should allocate more funding for the development, refinement and standardization of metrics and tools for measuring scientific changes. The gradual success of the largest international bibliographical databases made available by commercial providers as well as the OA movement in the public sector, presents encouraging examples of what can be achieved to reach such ambitious goals within a manageable timespan. For example, significant stepwise progress is made in Europe on the developing advanced metrics of open access publications (LERU, 2024). Existing international organizations and statistical agencies - such as the OECD in Paris with 60 years of experience in operating information platforms of country-level R&D statistics, and the UNESCO-produced series of *Science Reports* since 1993, could play an important role in these processes.

In conclusion, we are cautiously optimistic that the measurability of large-scale changes in the global science system can be significantly improved. By building on the currently available data infrastructures and information sources, and by embracing multi-stakeholder collaboration and innovative analytical frameworks, the science system itself has the capacity to integrate a broader range of high-quality, objective information on its own performance and its evolutionary trajectories. Authoritative global surveys, advanced metrics, smart indicators and AI-supported analytics would lead to more convincing empirical evidence, relevant new insights, better informed decision-making on how to drive and support scientific advancement. Sixty years later, DdSP would no doubt be delighted to see such progress.

## Literature references